\documentclass[twocolumn,twoside,slac_two]{revtex4}
\usepackage{graphicx}
\usepackage{fancyhdr}
\begin{document}

\title{Charged Lepton-Flavor Violation in Beyond-Standard Models}

%

\author{Junji Hisano}
\affiliation{Institute for Cosmic Ray Research (ICRR), 
University of Tokyo, Kashiwa, Chiba 277-8582, Japan}

\begin{abstract}
  I discuss charged lepton-flavor violation in physics beyond the
  standard model and review topics related to it. 
\end{abstract}

\maketitle

\thispagestyle{fancy}


\section{Introduction}

It is found from discovery of the neutrino oscillation that the
lepton-flavor symmetries are not exact in nature, while the charged
lepton-flavor violation (cLFV) has not been observed yet.  The cLFV is
suppressed even if the tiny neutrino masses observed in the
experiments are introduced in the standard model (SM). The cLFV
processes are proportional to the fourth power of the neutrino masses
due to the GIM mechanism, similar to the quark sector.  The predicted
branching ratio of $\mu\rightarrow e \gamma$ is smaller than
$10^{-54}$. However, the lepton-flavor symmetries are accidental in
the SM, and it is a big mystery in particle physics why the cLFV
processes are not discovered, when considering beyond-standard models
(BSMs) \cite{Raidal:2008jk}.

Searches for cLFV processes have long history. Search for
$\mu\rightarrow e \gamma$ was performed soon after muons were
discovered, and it was found that muons are not an excited state of
electrons. On 60's, the two-neutrino hypothesis, in which the lepton
flavor conservations were introduced, was proposed to suppress
$\mu\rightarrow e \gamma$.  The tau-lepton flavor conservation has
been also tested after tau lepton was discovered on 75'. However, on
98', the neutrino oscillation was discovered at SuperKamiokande
experiments. 

Now the bounds on the cLFV processes are significantly
improved by efforts of experimentalists.  The current experimental
bounds on the representative cLFV processes of muon 
are given as follows,
\begin{eqnarray}
{\rm Br}(\mu\rightarrow e \gamma) &<&1.2 \times 10^{-11} ~\mbox{\cite{Brooks:1999pu}}, \label{mueg}\\
{\rm Br}(\mu\rightarrow eee)  &<&1.0 \times 10^{-12}~\mbox{\cite{Bellgardt:1987du}},\\
{\rm R}(\mu\rightarrow e;{\rm Ti})  &<& 6.1 \times 10^{-13}~\mbox{\cite{Dohmen:1993mp}},
\label{mueti} \\
{\rm R}(\mu\rightarrow e;{\rm Au}) &<& 7.0 \times 10^{-13}~\mbox{\cite{Bertl:2006up}}.
\label{mueau}
\end{eqnarray}
$\mu$-$e$ conversion rates in nuclei, given in Eqs.~\ref{mueti}
and \ref{mueau}, are normalized by the muon capture rates.  Those for
tau-lepton's processes come from the Belle and Babar experiments, and
they are\footnote{
  The bounds on Eqs.~\ref{taulfvmu} and \ref{taulfve} are combined
  results of Belle \cite{Abe:2006sf} and Babar
  \cite{Aubert:2005ye,Aubert:2005wa}.
} 
\begin{eqnarray}
{\rm Br}(\tau\rightarrow \mu \gamma) &<&1.6 \times 10^{-8} ~\mbox{\cite{Banerjee:2007rj}}, \label{taulfvmu}\\
{\rm Br}(\tau\rightarrow e \gamma) &<&9.4 \times 10^{-8}~\mbox{\cite{Banerjee:2007rj}},
\label{taulfve}\\
{\rm Br}(\tau\rightarrow 3l ) &<&\sim 10^{-8}~\mbox{\cite{Banerjee:2007rj}},\\
{\rm Br}(\tau\rightarrow l+{\rm hadron(s)} ) &<&\sim 10^{-8}~\mbox{\cite{Banerjee:2007rj}}.
\end{eqnarray}

A new experiment for $\mu \rightarrow e \gamma$ search, MEG, has
been started at PSI this year \cite{meg}. It is argued \cite{tmori}
that a successful physics run for 5-6 months reaches to $(7$-$8)\times
10^{-13}$, which is improvement of more than 10 compared with the
current bound derived by the MEGA experiment on 98'
\cite{Brooks:1999pu}, and that a goal of the first phase of the
experiment is $(1$-$2)\times 10^{-13}$ after another two years. Further
improvement down to $\sim 10^{-14}$ in the second phase may be
possible after some upgrades.

The signal in $\mu$-$e$ conversion experiments is monochromatic
electron with energy $m_\mu-E_{\rm bound}$ ($E_{\rm bound}$ bounding
energy), and then the backgrounds (BGs) are quite suppressed. While
the conversion rates are suppressed by $10^{-(2-3)}$ compared with $Br(
\mu\rightarrow e \gamma)$ in typical BSMs such as supersymmetric
models, the high sensitivities to the BSMs are expected.  Two
experiments for $\mu$-$e$ conversion searches with sensitivities
$10^{-(16-17)}$, Mu2e in Fermilab \cite{mu2e} and COMET in J-parc
\cite{comet}, are planed.  Furthermore, the PRISM/PRIME experiment, in
which very intensive pulsed beam is produced by the FFAG muon storage
ring, is also planed. It has ultimate sensitivity as $10^{-(18-19)}$
\cite{comet}.

The searches for cLFV in the tau decay will be continued by the Belle
experiment; however, further improvement requires higher luminosity
${\cal L}$.  Now, two experiments, the Super KEKB
\cite{Akeroyd:2004mj} and the Super flavor factory
\cite{Giorgi:2006qj}, are planed. The processes $\tau\rightarrow \mu
\gamma$ and $\tau\rightarrow e \gamma$ already suffer from irreducible
BGs from $e^-e^+\rightarrow \tau^+\tau^-\gamma$, Improvements of the
sensitivities are scaled as $\sqrt{\cal L}$ without reduction of the
BGs, and the reaches are argued $10^{-(8-9)}$.  Other cLFV processes
in the tau decay are almost BG-free, and the reaches are argued
$10^{-(9-10)}$.
        
\section{CLFV in BSMs}

Following are effective operators with cLFV of muon up to $D=6$,
\begin{eqnarray}
{\cal L}
=\frac{m_\mu}{\Lambda^2}
\bar{e}\sigma^{\mu\nu}F_{\mu\nu}
\mu
+
\frac{1}{\Lambda_F^2}
\bar{e}\mu\bar{e} e
+
\frac{1}{\Lambda_F^2}
\bar{e}\mu\bar{q} q.
\label{efflfv}
\end{eqnarray}
The first term is for $\mu\rightarrow e \gamma$, and the second and
third ones are for $\mu\rightarrow 3e$ and $\mu$-$e$ conversion in nuclei,
respectively. Here, other LFV terms with different tensor structures
are omitted for simplicity. Roughly speaking, the branching ratios or
conversion rates are $\sim (m_W/\Lambda)^4$ or $ (m_W/\Lambda_F)^4$.
Thus, $\Lambda$ and $\Lambda_F$ should be larger than $\sim
10^{(5-6)}$~GeV from the experimental bounds. This implies that the
cLFV searches have quite sensitivities to the BSMs.

We have several motivations to consider the BSMs at TeV scale. The
naturalness argument for the Higgs boson mass terms is one of them.
The dark matter in the universe may be related to physics at TeV
scale. In addition, some people consider origin of the neutrino masses
at TeV scale.
 
Why does nature hide clues of the BSMs from FCNC processes including
cLFV processes? Some of BSMs, such as the two- or multi-Higgs doublet
models and the left-right symmetric models, introduce new Higgs and/or
gauge bosons at TeV scale. Some models, such as supersymmetric models,
extra dimension models, and the little Higgs model with T parity,
introduce partners of leptons and quarks. Those new fields
introduce new sources of FCNCs. While FCNCs in the SM are suppressed
by small quark/lepton masses or small mixing angle due to the GIM
mechanism, the suppression is not necessarily automatic in the BSMs.

First, let us consider the cLFV processes in the supersymmetric
standard model (SUSY SM), since it is the leading candidate among the
BSMs and also a prototype of the BSMs. Supersymmetry is a symmetry
between bosons and fermions, and superpartners for each particles in
the SM are introduced to the SUSY SM.

Supersymmetry is not exact in nature. We need to introduce
SUSY-breaking mass terms for SUSY particles, which have not yet been
discovered. The SUSY-breaking terms are new sources of the flavor
violation, since the squark and slepton mass matrices are not
necessarily simultaneously diagonalized with those of quarks and
leptons.\footnote{
The R parity is assumed here. When it is violating, new LFV sources
are introduced in the SUSY SM. See Ref.~\cite{deGouvea:2000cf} for the
detail discussion.
}
This leads to so-called the SUSY flavor problem.  We will discuss this
problem from a viewpoint of cLFV in the following.

In the SUSY SM $\mu \rightarrow e \gamma$ is generated by one-loop
diagrams, and the branching ratio is approximately given as
\begin{eqnarray}
{\rm Br}(\mu\rightarrow e \gamma) \sim \frac{\alpha}{4\pi}
\left(\frac{m_W}{m_{SUSY}}\right)^4
 \sin^2\theta_{\tilde{e}\tilde{\mu}}
\left(\frac{\Delta m_{\tilde{l}}^2}{m^2_{SUSY}}\right)^2,
\nonumber\\
\end{eqnarray}
where $m_{SUSY}$ the SUSY breaking scale,
$\sin\theta_{\tilde{e}\tilde{\mu}}$ the slepton mixing angle, ${\Delta
  m_{\tilde{l}}^2}$ is the slepton mass square difference.

While the branching ratio of $\mu\rightarrow e \gamma$ is suppressed
by one-loop factor, we need much more suppression in order to make it
below the experimental bound. Three directions are proposed. First is
the universal scalar mass hypothesis, in which the squarks and
sleptons with common quantum numbers are degenerate in mass (${\Delta
  m_{\tilde{l}}^2}\ll m_{SUSY}^2$). Many models are constructed
following this direction; the gravity mediation \cite{Nilles:1983ge},
the gauge mediation \cite{gaugemed}, the gaugino mediation
\cite{ginom}, and the anomaly mediation \cite{ams}.  Second is the
alignment hypothesis ($\sin\theta_{\tilde{e}\tilde{\mu}}\ll 1$).  It
is assumed that squark and slepton mass matrices can be diagonalized
in the same basis as those of quarks and leptons due to some flavor
symmetries or some mechanism \cite{Nir:1993mx}.  Third is the
decoupling hypothesis ($m_{SUSY}\gg m_W$).  Squarks and sleptons in
the first and second generations are so heavy ($O(10^{4-5})$~GeV) that
the flavor violation in the first and second generations are
suppressed \cite{effsusy}. 

In these three hypothesis, the cLFV processes are suppressed.
However, the improvements of the experimental sensitivities 
may probe the origin of the SUSY breaking terms and also physics
beyond the SUSY SM. 

\begin{figure}[h]
\centering
\includegraphics[width=60mm]{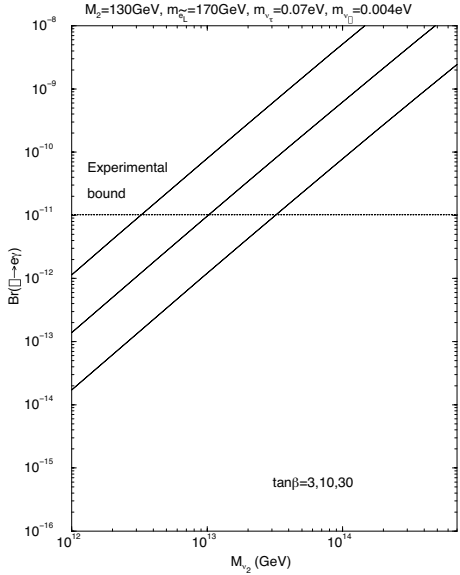}
\caption{${\rm Br}(\mu\rightarrow e \gamma)$ in the SUSY seesaw
  model. This figure comes from Ref.~\cite{Hisano:1998fj}.
} \label{example_figure}
\end{figure}

In the universal scalar mass hypothesis, if some physics has LFV
interactions below the SUSY-breaking mediation scale, the LFV slepton
mass terms are induced radiatively by the renormalization-group effect
\cite{Hall:1985dx}. In this case the LFV mass terms are not suppressed
by powers of the energy scale for the LFV interactions. The seesaw
mechanism and the GUTs are nowadays ones of attractive models from the
phenomenological and theoretical points of view. In these models, LFV
Yukawa interactions are introduced. Thus, if the SUSY breaking
mediation scale is higher than the GUT \cite{Barbieri:1994pv} or the
right-handed neutrino mass scale \cite{Borzumati:1986qx}, sizable LFV
processes might be predicted. In Fig.~\ref{example_figure}, ${\rm
  Br}(\mu\rightarrow e \gamma)$ in the SUSY seesaw model is shown.
The observed large neutrino mixing angles enhances the cLFV processes
of muon and tau lepton \cite{Hisano:1997tc,Hisano:1998fj}. Various
studies are performed under the universal scalar mass hypothesis.  See
references in Ref.~\cite{Raidal:2008jk}.

\begin{figure}[h]
\centering
\includegraphics[width=60mm]{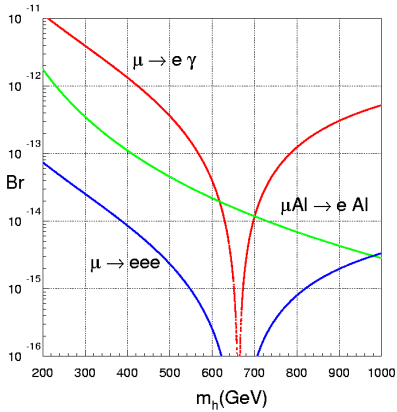}
\caption{Branching ratios for cLFV processes of muon induced
by the Higgs boson exchange in the SUSY SM.
This figure comes from Ref.~\cite{Paradisi:2006jp}.
} \label{higgsmed}
\end{figure}

In the decoupling hypothesis, squarks and sleptons in the third
generation may have large flavor violation. In addition, even when
SUSY particle masses larger than $O(1$-$10)$~TeV, the Higgs boson
exchange may give sizable contributions to the cLFV processes
\cite{Babu:2002et}. The SUSY SM has two doublet Higgs bosons ($H_1$ and
$H_2$), and their effective couplings for leptons are
\begin{eqnarray}
-{\cal L}_Y= \bar{e}_{Ri} Y_{li} L_i H_1+
\bar{e}_{Ri} \Delta_{ij} L_j H_2^\dagger+h.c.,
\end{eqnarray}
where $Y_{li}$ is the tree-level Yukawa coupling and $\Delta_{ij}$ is
non-holomorphic correction generated at one-loop level. $\Delta_{ij}$
is not suppressed by the SUSY breaking scale, in addition to
enhancement of $\tan\beta$. Then, when the SUSY particles are heavy
enough, the Higgs boson exchange dominates in the cLFV processes.
Fig.~\ref{higgsmed} shows the branching ratios for cLFV processes of
muon, induced by the Higgs boson exchange in the SUSY SM.  This figure
comes from Ref.~\cite{Paradisi:2006jp}. $\mu\rightarrow e \gamma$ is
induced by the Barr-Zee type loop diagrams, while $\mu\rightarrow 3e$
and $\mu\rightarrow e $ conversion in nuclei are Higgs-boson exchange
processes effectively at tree-level. The Higgs boson exchange
contributions to the cLFV processes of tau lepton are also discussed
in Ref.~\cite{higgstau}.

If the alignment between lepton and slepton masses is not complete in
the third hypothesis, the cLFV processes are predicted. Let us show
an example with $U(1)\times U(1)$ flavor symmetries, which is  given in
Ref.~\cite{Feng:2007ke}.  The charge assignments
of the right- and left-handed leptons are
\begin{eqnarray}
\bar{E}_1(1,0),~\bar{E}_2(1,-2),~\bar{E}_3(0,-3),\nonumber\\
L_1(4,0),~L_2(2,2),~L_3(0,4),\nonumber
\end{eqnarray}
and the flavor symmetries are broken by VEVs of flavon fields,
$\phi_1(-1,0)$ and $\phi_2(0,-1)$. In this case, the charged
lepton mass matrix is
\begin{eqnarray}
(m_l)\sim\lambda \left(
\begin{array}{ccc}
\lambda^4&0&0\\
\lambda^4&\lambda^2&0\\
\lambda^4&\lambda^2&1
\end{array}
\right),
\end{eqnarray}
and the hierarchical structure of lepton masses is explained with
$\lambda \sim 0.1-0.2$.  In this setup, the slepton mass matrices are
aligned, and the left-handed and right-handed slepton mixings are
suppressed as $\theta_{\tilde{\mu}_L\tilde{e}_L}\sim \lambda^4$ and
$\theta_{\tilde{\mu}_R\tilde{e}_R}\sim \lambda^2$. However, these
mixings are marginal to the bounds from $\mu\rightarrow e\gamma$. 

The SUSY flavor problem is one of the guidelines to construct
realistic SUSY SMs. As shown above, in the proposed ideas and models
to suppress the FCNC processes, the cLFV processes are not necessarily
suppressed, and the on-going and planed experiments cover the
predictions.

The cLFV processes also have good sensitivities to non-SUSY models at
TeV scale as expected.  The BSMs at TeV scale are severely constrained
from the cLFV processes, unless some mechanism works to suppress them. In
the following, some concrete examples are reviewed.

First is the little Higgs model \cite{ArkaniHamed:2001ca} with T
parity \cite{Cheng:2003ju}. In the little Higgs model, the Higgs boson
is a pseudo Nambu-Goldstone boson for symmetry breaking, and quadratic
divergence in the radiative correction to the Higgs boson mass term is
cancelled by heavy particles at one-loop level. The T parity is
introduced for the heavy particles not to contribute to the
electroweak observables at tree level. This extension has a bonus.
The lightest T-odd particle is stable and a candidate of the dark
matter in the universe.

\begin{figure}[h]
\centering
\includegraphics[width=60mm]{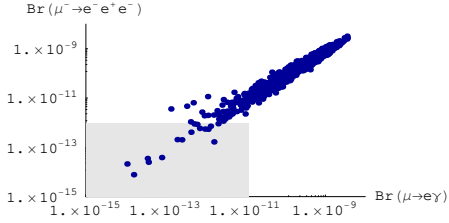}
\caption{${\rm Br}(\mu\rightarrow e \gamma) $
and ${\rm Br}(\mu\rightarrow 3e)$ in the little Higgs model with T parity.
Here, the mirror leptons masses are taken from 300~GeV and 1.5~TeV.
This figure comes from Ref.~\cite{Blanke:2007db}.
} \label{littlehiggs}
\end{figure}

In the little Higgs model with T parity, SU(2)$_L$ doublet mirror
leptons with T parity odd are introduced. The mirror leptons have
coupling with leptons and the heavy gauge bosons, which is
lepton-flavor violating. The cLFV processes are generated at one-loop
level, similar to the SUSY SM. In Fig.~\ref{littlehiggs}, ${\rm
  Br}(\mu\rightarrow e \gamma) $ and ${\rm Br}(\mu\rightarrow 3e)$ are
shown in this model. It is shown in Ref.~\cite{Blanke:2007db} that
the accidental cancellation reduces ${\rm Br}(\mu\rightarrow e \gamma)$
and two processes have comparable branching ratios. Then, this model
still is viable even the mirror leptons are around 1~TeV.

\begin{figure}[h]
\centering
\includegraphics[width=50mm,angle=90]{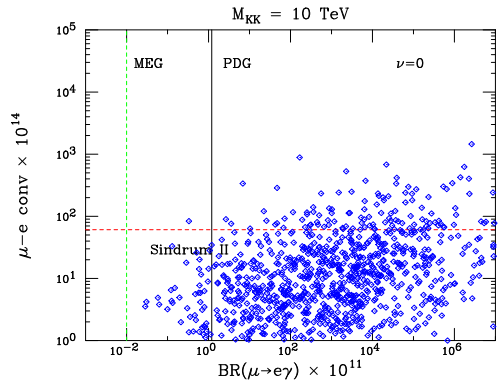}
\caption{ ${\rm Br}(\mu\rightarrow e \gamma) $ and ${\rm
    R}(\mu\rightarrow e;{\rm Ti})$ in the SM on the Randall-Sundrum
  background.  Here, the Kaluza-Klein scale is 10~TeV.  This figure
  comes from Ref.~\cite{Agashe:2006iy}.  } \label{rs}
\end{figure}

Next is the SM on the Randall-Sundrum (RS) background. The RS geometry
is known as a solution of the hierarchy problem \cite{Randall:1999ee}.
In addition, when the SM fermions and gauge bosons propagate in the
full five-dimensional space, the fermion mass hierarchy is also
explained even from the ``anarchic'' structure
\cite{Gherghetta:2000qt}. In this model the Kaluza-Klein (KK)
particles have the LFV interaction. In Fig.~\ref{rs} ${\rm
  Br}(\mu\rightarrow e \gamma) $ and ${\rm R}(\mu\rightarrow e;{\rm
  Ti})$ are shown for the KK scale 10~TeV. The current experimental
bounds give constraints on this model.  
While $\mu\rightarrow e
\gamma$ is a one-loop process, the $\mu\rightarrow e$ conversion is
generated at tree level.

Finally, we discuss the relation of the cLFV processes and the
neutrino masses.  The seesaw mechanism is the most promising in the
candidates of the neutrino masses. In many models the mechanism is
realized at much higher energy scale than the TeV scale. In such
models we cannot expect some direct relations between the cLFV
processes and the origin of the neutrino masses. Exceptions are the
SUSY seesaw models \cite{susyseesaw1,susyseesaw2}; however, the
relations are indirect. On the other hand, there are attempts to
construct models of the neutrino mass origin at TeV scale.

\begin{figure}[h]
\centering
\includegraphics[width=60mm]{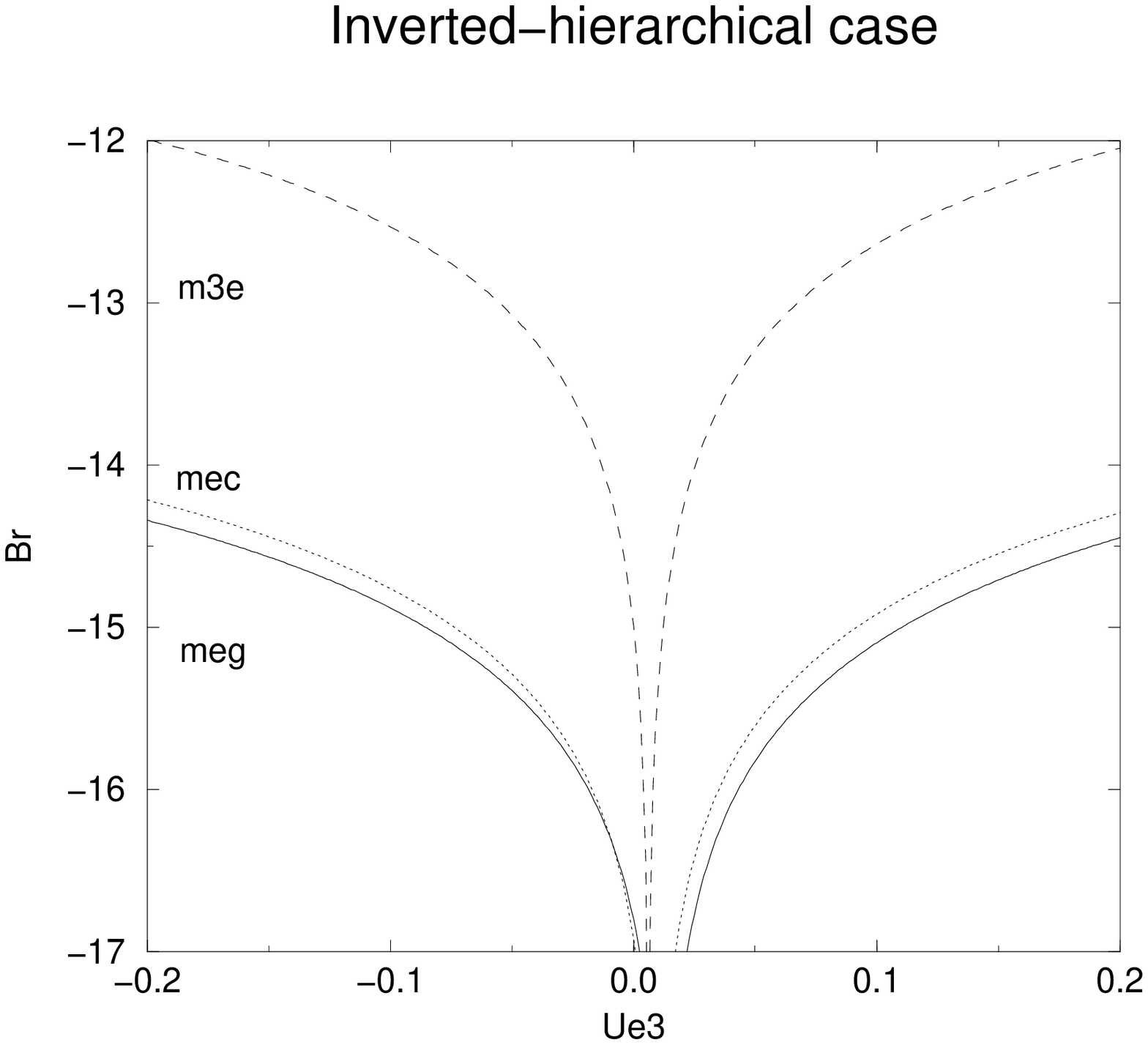}
\includegraphics[width=60mm]{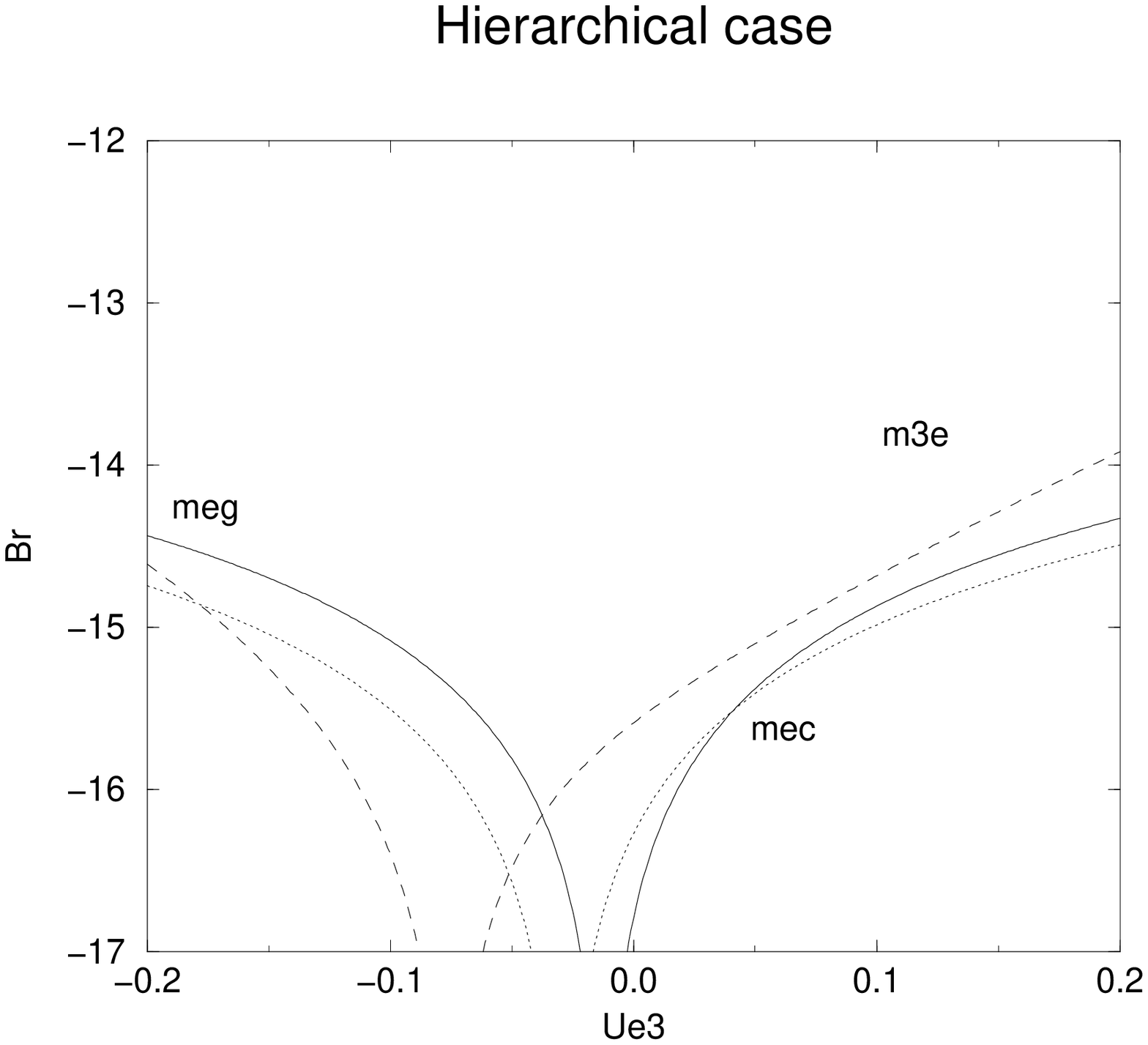}
\caption{Branching ratios for cLFV processes of muon 
in the triplet Higgs model. Here, neutrino mass spectrum
is inverted-hierarchical (hierarchical) in the upper (lower) figure.
These figures come from Ref.~\cite{Kakizaki:2003jk}
} \label{triplet}
\end{figure}

One of the models is the triplet Higgs model (type-II seesaw model)
\cite{Schechter:1980gr}.  In this model, the triplet Higgs boson with
mass at TeV scale has sizable lepton-flavor violating coupling, and
the coupling is directly linked to the neutrino mass matrix. In
Fig.~\ref{triplet} the branching ratios for the cLFV processes of muon
in this model are shown. The pattern depends on the neutrino mass
structure \cite{Kakizaki:2003jk}.

\section{Correlations}

When the cLFV processes are discovered, it would be important to take
correlations among various processes in order to unveil the origin of
the lepton-flavor violation and the realistic BSM. 

The first correlation is among the cLFV processes with common flavor
transition. The correlation discriminates BSMs at TeV scale since the
pattern of correlation depends on models. In Table~\ref{ratioratio},
the ratios of the cLFV rates of muon are shown in three models; the
SUSY SM in which the dipole moment contribution to the cLFV processes
is dominant, the SUSY SM in which the SUSY particles are so heavy that
the Higgs boson exchange dominates, and the little Higgs model with T
parity. While the cLFV processes are radiatively generated in those
models, the pattern of the ratio of the cLFV rates are quite different.
Correlations among the cLFV processes of $\tau$ and among the
$\mu$-$e$ conversions in various nuclei \cite{Kitano:2002mt} are also
useful to discriminate models. 

\begin{table}[h]
\begin{center}
  \caption{Ratios of the cLFV rates of muon in three models. This
    table comes from \cite{Blanke:2007db}.  }
\begin{tabular}{|c||c|c|}
\hline 
& $\frac{{\rm Br}(\mu\rightarrow 3e)}{{\rm Br}(\mu\rightarrow e\gamma)}$
& $\frac{{\rm R}(\mu\rightarrow e;{\rm Ti})}{{\rm Br}(\mu\rightarrow e\gamma)}$
\\
\hline 
SUSY SM (dipole) &$\sim 6\times 10^{-3}$&$\sim 5\times 10^{-3}$ \\
SUSY SM (Higgs) &$\sim 6\times 10^{-3}$&$0.08$-$0.15$ \\
Little Higgs with T parity   &0.4-2.5&$10^{-2}$-$10^2$ \\
\hline
\end{tabular}
\label{ratioratio}
\end{center}
\end{table}

The second correlation is among the cLFV processes with different
flavor transitions. In some models the cLFV processes are related to
the neutrino masses. In the triplet Higgs model, the interaction of
the triplet Higgs boson with leptons is directly linked to the
neutrino mass matrix $(m_\nu)$, as mentioned above. The model is one of 
realizations of the minimal flavor violation hypothesis in the lepton
sector \cite{Cirigliano:2005ck}. As the result, the ratios of the cLFV
rates of tau lepton and muon are given as
\begin{eqnarray}
\frac{{\rm Br}(\tau\rightarrow \mu\gamma)}{{\rm Br}(\mu\rightarrow e\gamma)} 
\simeq0.17\times
\left(
\frac{(m_\nu m_\nu^\dagger)_{\tau\mu}}
{(m_\nu m_\nu^\dagger)_{\tau e}}
\right)^2\sim 300\nonumber\\
\frac{{\rm Br}(\tau\rightarrow e\gamma)}{{\rm Br}(\mu\rightarrow e\gamma)} 
\simeq0.17\times
\left(
\frac{(m_\nu m_\nu^\dagger)_{\tau e}}
{(m_\nu m_\nu^\dagger)_{\tau e}}
\right)^2\sim 0.2.
\label{minimalflavor}
\end{eqnarray}

In the SUSY seesaw models, the cLFV processes are indirectly related
to the origin of the neutrino masses even if the energy scale is high.
Under the universal scalar mass hypothesis, the cLFV processes give
information of the type-I SUSY seesaw model, which is independent of the
neutrino mass from a viewpoint of the reconstruction of the model
\cite{susyseesaw2}. In the type-II SUSY seesaw model, the relation in
Eq.~\ref{minimalflavor} is predicted again \cite{susyseesaw2}.

The third correlation is between the hadronic and leptonic FCNC
processes. In GUTs, quarks and leptons are unified so that the cLFV
processes are correlated with hadronic FCNC processes
\cite{Hisano:2003bd}. Let us consider the SUSY SU(5) GUT with right-handed
neutrinos. In this model, lepton doublets and right-handed down-type
quarks are embedded in common SU(5) multiplets. Then, neutrino Yukawa
coupling affects both the right-handed sdown and left-handed slepton
mass matrices \cite{Moroi:2000mr}.

\begin{figure}[h]
\centering
\includegraphics[width=60mm]{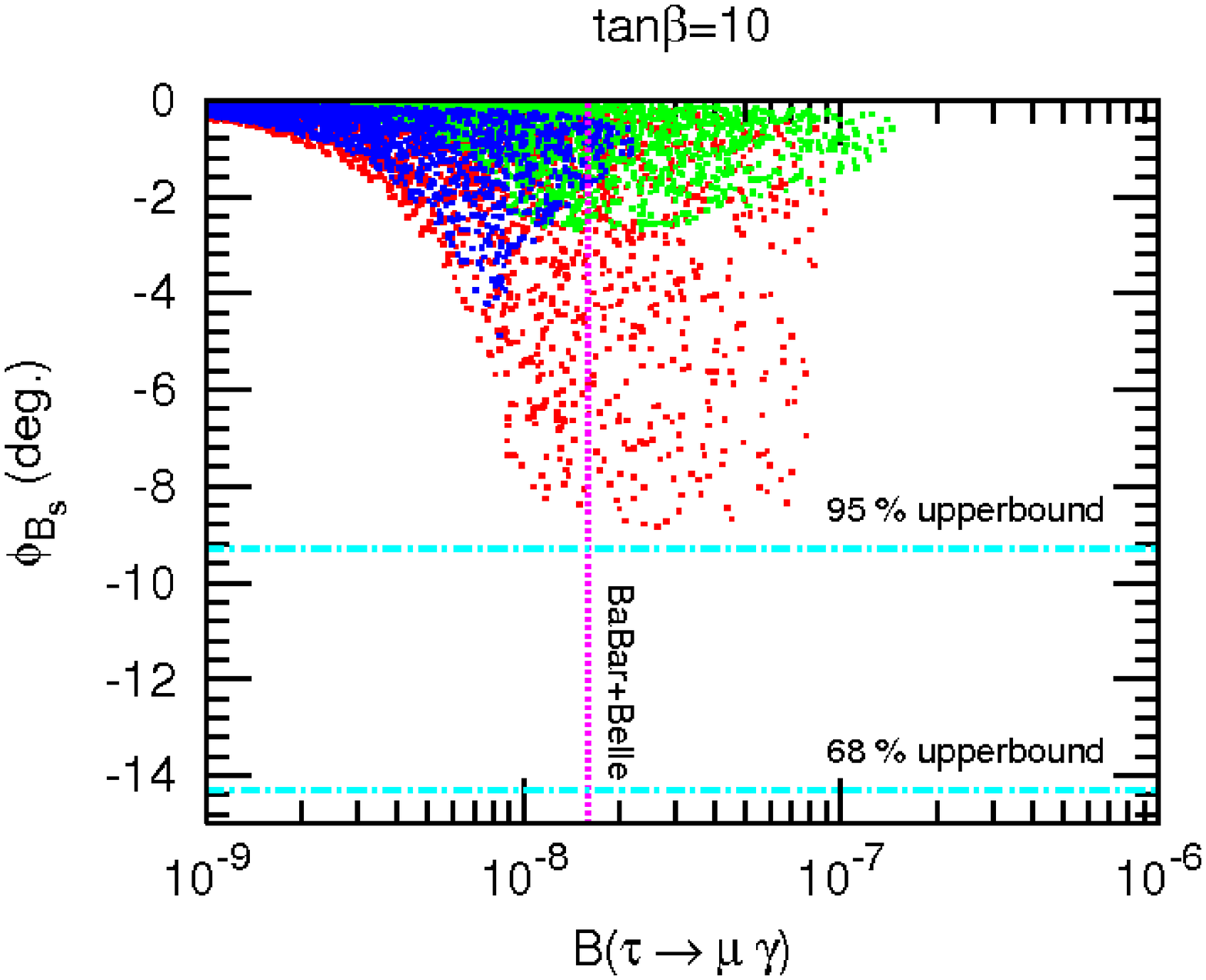}
\includegraphics[width=60mm]{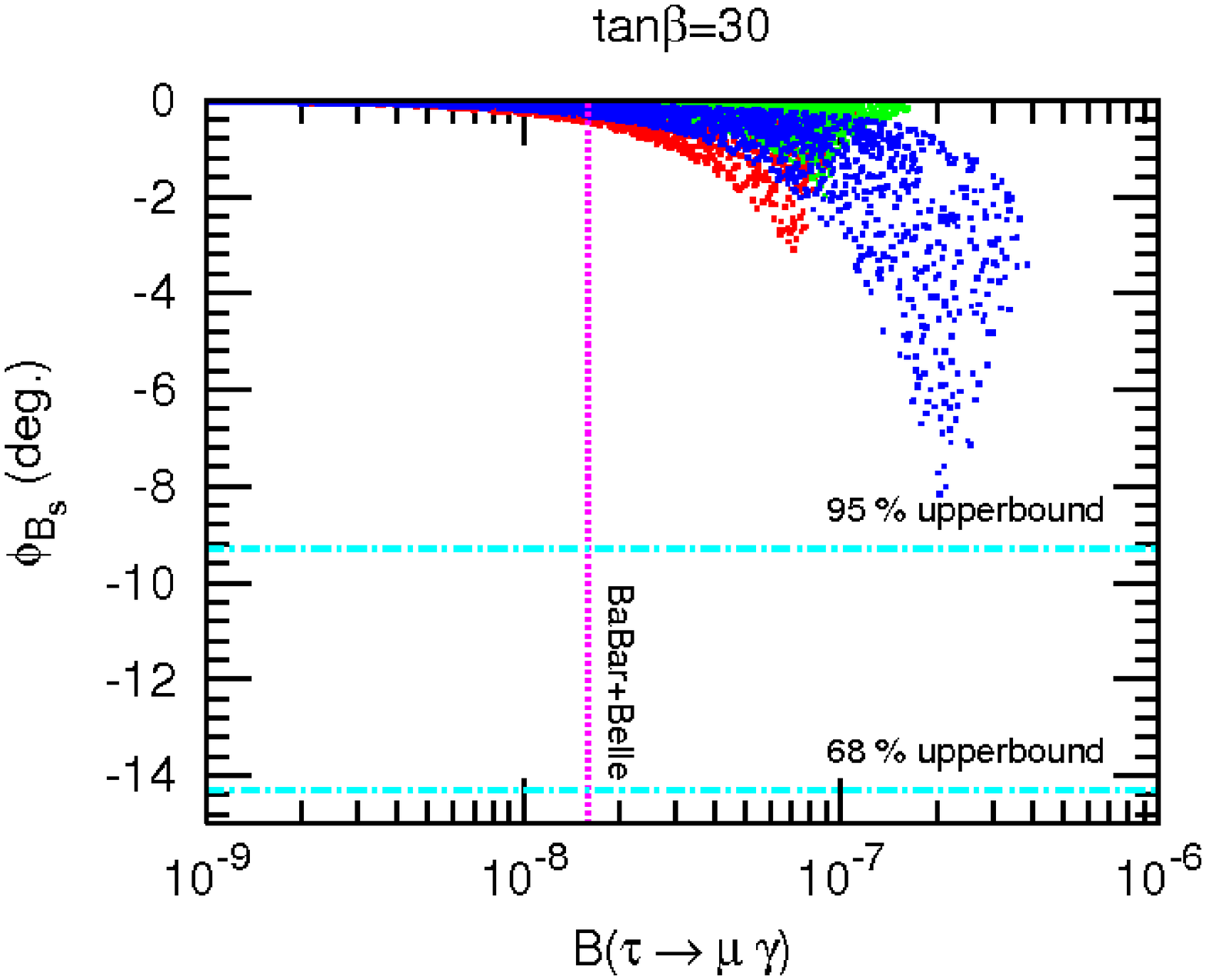}
\caption{ Correlation between CP phase of the $B_s$ mixing amplitude,
  $\phi_{B_s}$, and ${\rm Br}(\tau\rightarrow \mu \gamma)$ in the SUSY
  SU(5) GUT with right-handed neutrinos. Here, $\tan\beta=10$ and 30,
  and we show the region for $\phi_{B_s}$, derived by the {\bf Ut}{\it
    fit} collaboration.  These figures come from
  Ref.~\cite{Hisano:2008df} } \label{susygut}.
\end{figure}

In Fig.~\ref{susygut} we show the correlation between CP phase of the
$B_s$ mixing amplitude, $\phi_{B_s}$, and ${\rm Br}(\tau\rightarrow
\mu \gamma)$ in the SUSY SU(5) GUT with right-handed neutrinos.  The
CP violation in the $B_s$ mixing is suppressed in the SM, and then it
is also sensitive to the BSMs. The phase $\phi_{B_s}$ is defined to be
zero in the SM.  It is recently announced by the {\bf Ut}{\it fit}
collaboration \cite{Bona:2008jn} that $\phi_{B_s}$ deviates more than
$3 \sigma$ from the SM prediction. In the figure, we show the region
for $\phi_{B_s}$. The deviation of $\phi_{B_s}$ is constrained from
null result of $\tau\rightarrow\mu \gamma$ search in this model, and
the $95\%$ probability region derived by the {\bf Ut}{\it fit}
collaboration is marginal in this model. When the deviation of
$\phi_{B_s}$ is established, search for $\tau\rightarrow \mu \gamma$
would be an important test of the SUSY SU(5) GUTs.
 
When the flavor-violating mass terms for the right-handed squarks are
non-vanishing, the hadronic EDMs are generated at one-loop
\cite{Dimopoulos:1994gj} and two-loop levels \cite{Hisano:2007cz}.
The non-zero $(m_{\tilde{d}_R}^2)_{23}$ generates the strange-quark
chromoelectric dipole moment, which contributes to the hadronic EDMs
\cite{Hisano:2003iw}. Thus, the correlation between the hadronic EDMs
and the cLFV processes would be the tests of the SUSY GUTs. But, this
program still has difficulties in precisions of the hadronic EDM
evaluation \cite{Narison:2008jp}, and the further improvements would
be required for it.

\section{Summary}

In this talk, I discussed charged lepton-flavor violation in physics
beyond the standard model and reviewed topics related to it.  The cLFV
processes are accidentally suppressed by the GIM mechanism in the SM
even after tiny neutrino masses are introduced. On the other hand, the
suppression is not necessarily automatic in physics beyond the SM.
Studies of cLFVs probe BSMs, hidden flavor symmetries, and underlying
flavor structures.

Current bounds on cLFVs give constraints on physics even around
O($10^{(5-6)}$) GeV. In practical BSMs, cLFVs are suppressed by
loop-factors or small flavor-mixing, or accidental cancellation. Thus,
coming MEG experiment, and planed experiments, Mu2e, COMET and
PRISM/PRIME, will cover interesting regions in various BSMs. Since the
cLFVs are pieces of puzzles in the BSMs, it is important to stress that
taking various correlations are useful to solve the puzzles.

\begin{acknowledgments}
  The work of JH was also supported in part by the Grant-in-Aid for
  Science Research, Japan Society for the Promotion of Science
  (No.~20244037 and No.~2054252).
\end{acknowledgments}

\bigskip 

\end{document}